A universal model for the formation energy prediction of inorganic compounds

Yingzong Liang[1, 2], Mingwei Chen[1], Yanan Wang[1, 2], Huaxian Jia[2], Tenglong Lu[2], Fankai Xie[2], Sheng Meng[1, 2*], Miao Liu[1, 2, 3*]

[1]Songshan Lake Materials Laboratory, Dongguan, Guangdong, 523808, China
[2]Beijing National Laboratory for Condensed Matter Physics, Institute of Physics, Chinese Academy of Sciences, Beijing, 100190, China
[3]Center of Materials Science and Optoelectronics Engineering, University of Chinese Academy of Sciences, Beijing, 100049 P. R. China

*Corresponding author: mliu@iphy.ac.cn, smeng@iphy.ac.cn

## ABSTRACT

Harnessing the recent advance in data science and materials science, it is feasible today to build predictive models for materials properties. In this study, we employ the data of high-throughput quantum mechanics calculations based on 170,714 inorganic crystalline compounds to train a machine learning model for formation energy prediction. Different from the previous work, our model reaches a fairly good predictive ability ($R^2$=0.982 and MAE=0.07 eV·atom$^{-1}$, DenseNet model) and meanwhile can be universally applied to the large phase space of inorganic materials. The improvement comes from several effective structure-dependent descriptors that are proposed to take the information of electronegativity and structure into account. This model can provide a useful tool to search for new materials in a vast phase space in a fast and cost-effective manner.

## 1. Introduction

Formation energy, the energy to bind atoms together to form condensed matters, is one of the most important physical properties of a material. The formation energy represents the strength of the adhesion of atoms to each other within the material system, and many thermodynamic- and kinetic-related properties, such as stability [1] and synthesizability [2] of a compound, are directly associated with the magnitude of the formation energy. The value of the formation energy can be obtained from the first-principles calculations, which are relatively computation-intensive and expensive. Harnessing the advance of materials science computational databases as well as the knowledge of data science, the formation energy prediction can be achieved in an easier approach.

Previously, attempts have been made to quickly gauge the formation energy of inorganic materials system. E.g., Ward *et al.* [3] built a machine learning decision tree model by training with a dataset of 435,000 formation energies for inorganic compounds taken from the Open Quantum Materials Database (OQMD) [4], and achieved a mean absolute error (MAE) of 80 meV·atom$^{-1}$ in cross validation for predicting formation energies by employing structural descriptors from the Voronoi tessellation of the structure of crystalline. Cao *et al.* [5] utilized a convolution neural network (CNN) model and mixed Magpie descriptors [3] and Orbital-field matrix (OFM) descriptors [6] as input for predicting formation energy with more than 4,000 crystalline materials including transition metal binary alloys, lanthanide metal and transition metal binary alloys. The prediction performance of formation energy achieved MAE of 70 meV·atom$^{-1}$. Ye *et al.* [7] used the Pauling electronegativity [8, 9] and ionic radii [10] of the constituent species as the input descriptors with artificial neural networks (ANNs), and obtained a model with extremely low MAEs of 7–10 meV atom$^{-1}$ and 20–34 meV·atom$^{-1}$ in predicting the formation energies of garnets and perovskites, respectively. Xie and Grossman [11] proposed a generalized graph convolutional neural networks (CGCNN) model for material property predictions, and their model for formation energy predication reached the MAE of 39 meV·atom$^{-1}$ with the dataset of 28,046 samples taken from the Materials Project [12]. Li *et al.* [13] proposed to use deep neural network based transfer learning and a set of hybrid descriptors for perovskite formation energy prediction. The hybrid descriptors are composed of structural and elemental features as calculated via the Python Materials Genomics (*pymatgen*) library [14].

People have made effort on improving the predictive performance of models, and have showed that the formation energy of a given compound can be indeed roughly gauged from the Artificial Intelligence (AI) models. The previous works also hints on that the data quantity, data quality as well as the choice of descriptors of the models are crucial for this type of research. First, the larger the dataset the greater statistical power for pattern recognition [15]. As shown in Figure S1, the MAE of the formation energy prediction decreases monotonically with the size of the training set following a curve that decays with a power of the size of the training set. Second, the model should be built based on the physics of a system in order to yield an effective and accurate prediction. In our case of the formation energy prediction, the decisive factors of the atom-atom bonding strength should be embedded into the descriptor. It is also possible that the data can somewhat tell us the underlying physical mechanisms of materials from the statistic. In this work, we would like to use a large size computational materials science data to construct a universal AI model for formation energy prediction, and the model will also tell us what affects the formation energy by how much.

Generally speaking, it requires a large and high quality of dataset in order to extract the personae of materials. Several existing materials databases are tackling such data accumulation task, *e.g.*, Materials Project [12], AFLOWlib [16], Open Quantum Materials Database (OQMD) [4] and Novel Materials Discovery Laboratory (NOMAD) [17], all have done unprecedented job on generating huge amount of structural and property data for inorganic compounds [18] via the high-throughput quantum mechanics-based calculations. Those calculations, mostly the density functional theory (DFT), are relatively accurate and effective for obtaining the properties of a compounds via a *in silico* fashion, and the calculation power can scale out if one has enough computational resource and data-processing ability. In this work, we employ the dataset from the "atomly.net" [19], a recently developed DFT computational materials database with 170k compounds, all calculated with highly standardized procedures.

Unlike the previous work, the purpose of this work is to build a universally applicable model that can handle nearly any given inorganic crystalline compounds, hence it can help the materials community not only to estimate the energetic aspect of the compounds quickly, but also covers a large phase space. To make the predicting power of our model as good as possible, the descriptors are carefully designed and intensively tested. By analyzing the effectiveness of each descriptor on predicating the DFT-calculated formation energy, it is found that our proposed descriptors are fairly effective comparing with the previous descriptors, as our descriptors are physics-base.

From a broad view of the materials science, the ultimate goal of materials science is to understand the "structure-property" relation at the atomistic level. The recent advance on data-driven approach hints on that the Artificial Intelligence (AI) is likely to unlock such problem. Previously, AI has been employed to predict physical properties, such as formation energy [3, 5, 7, 11, 13], bandgap [20, 21], elastic constant [22], Seebeck coefficient [23], in existing publications. Legrain *et al.* [24] rapidly predicted vibrational free energies and entropies of crystalline materials by random forests and non-linear support vector machines. Lee *et al.* [25] used the PBE-band gap of 270 inorganic compounds to predict the values from $G_0W_0$ calculation and achieved the RMSE of 0.59 eV by nonlinear Support Vector Regression (SVR). Pilania *et al.* [26] predicted the HSE band gap of elpasolite ($A_2BB'X_6$-type) compounds calculated using a multi-fidelity co-kriging statistical learning framework and reported the accuracy of 0.1–0.2 eV on the validation set. Evans and Coudert [27] recently calculated elastic properties (bulk and shear moduli) for over one hundred silica zeolites by means of the DFT method, used them to train the prediction model with the gradient boosting regressor (GBR), and finally predicted 590,448 hypothetical zeolites. Zhang and Ling [28] used the band gaps simulated at the GGA-level to model the band gap of binary semiconductors from a small dataset. Artificial intelligence, and specially the machine learning as a branch of AI, is causing widespread disruption in almost every corner of materials science. AI or data-driven method has opened up a new avenue for us to extract information from materials science data, and it is possible that it would facilitate materials science significantly, hence the discovery of the new materials can be quicker and easier.

## 2. Results and discussion

Nine different ML models on prediction performance of $E_{form}$ were benchmarked by comparing their cross-validated coefficients of determination ($R^2$), root-mean-squared errors (RMSE), and mean absolute error (MAE) values of the test set. The optimized parameters of the eight conventional ML models are listed in Table S3. The effect of the new proposed SD descriptors was investigated by comparing changes in the $R^2$, RMSE, and MAE values before and after adding the SD descriptors. The top 20 important descriptors were screened out by using the RFR method, and the correlations between each descriptor pair were studied by the Pearson correlation coefficient.

To estimate the predictive performance of the ML models for $E_{form}$, the corresponding cross-validated $R^2$ values are illustrated in Figure 2. The horizontal and vertical axis are the DFT-calculated and predicted $E_{form}$, respectively. Note that the DN model gave the best performance with $R^2 = 0.982$, while the RR method has the lowest prediction accuracy with $R^2 = 0.858$, as shown in Figure 2a and 2i. It is proved that the four classical ML models, GBR, ABR, ETR, and SVR, can also achieve a very high prediction accuracy of $E_{form}$ with $R^2 > 0.95$ shown in Figure 2d–2h. For the two tree-based models, the ETR with $R^2 = 0.961$ is better than the RFR with $R^2 = 0.945$. As most classical boosting algorithms, both ABR (with extremely randomized trees as the weak learners) and GBR (with decision trees as the weak learners) can implemented very high $R^2 = 0.957$ and 0.952, respectively. The $R^2$ value of the KNNR model is close to 0.9 illustrated in Figure 2c. Compared Figure 2b to 2h, it is found that using the SVR model with the nonlinear rbf kernel function can get a better predication performance than that with the linear kernel function.

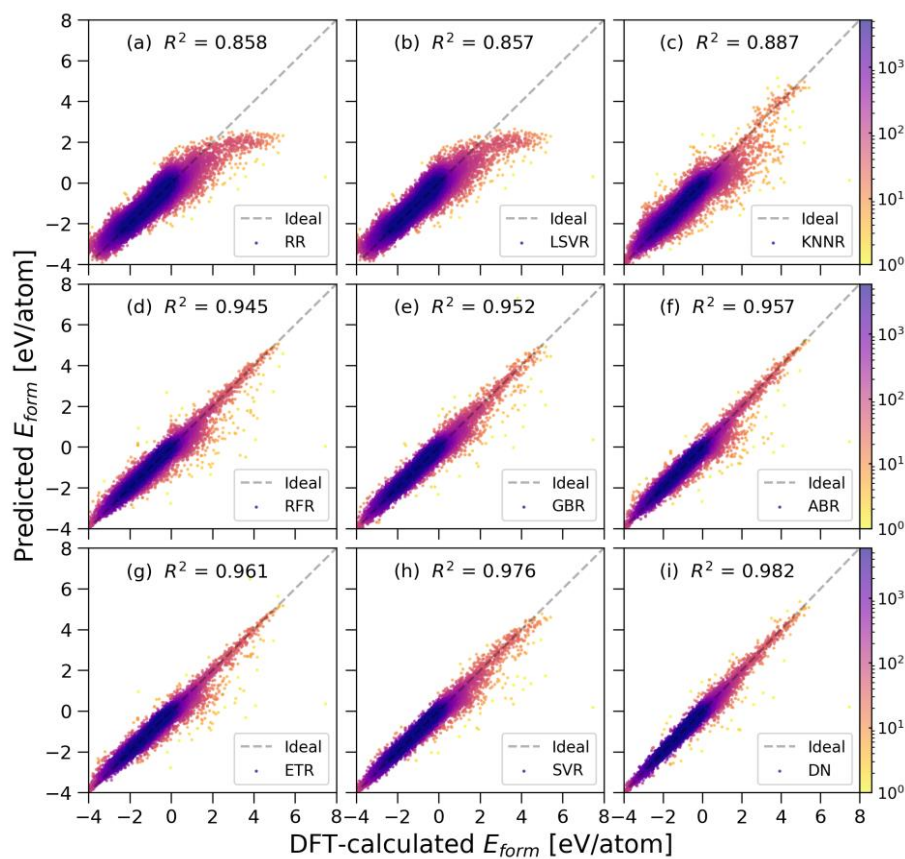

**Figure 2**. The DFT-calculated vs. predicted scatter plot of formation energy for the nine different ML methods with the new set of descriptors including both the CD and SD descriptors: (a) Ridge Regression (RR); (b) Linear Support Vector Regression (LSVR); (c) K-Nearest Neighbors Regression (KNNR); (d) Random Forest Regression (RFR); (e) GradientBoosting Regression (GBR); (f) AdaBoost Regression (ABR); (g) ExtraTrees Regression (ETR); (h) Support Vector Regression (SVR); (i) DenseNet (DN). The gray dash line represents the ideal curve y=x.

With the objective to estimate the effect of the new SD descriptors proposed in this study on predicting formation energy, we investigated the changes in evaluation metrics ($R^2$, RMSE, and MAE) of the ML models with and without the SD descriptors. As shown in Figure 3, the prediction performances of $E_{form}$ for all the nine ML models have been improved by increasing the SD descriptors, which indicate that the local structure information describes $E_{form}$ very well for the inorganic materials. Figure 3a depicts that the increase rate of $R^2$ values vary from 6.3% to 17.0% in combination of the SD descriptors with the CD ones. The SD descriptors dramatically improve the $R^2$ value of the ABR model for $E_{form}$ from 0.829 to 0.957. For DN model, which has the highest predictive power, the $R^2$ value increase from 0.885 to 0.982, which is 11.0% increase. The changes of RMSE and MAE values are shown in Figure 3b –3c, one can see that the decrease rates of RMSE and MAE values fluctuates in the range of [20.5%, 60.4%] and [10.0%, 50%], respectively. These results prove that the newly proposed structure-dependent descriptors, which is the electronegativity difference of neighboring atoms, can significantly improve the prediction performances of $E_{form}$ in the inorganic materials.

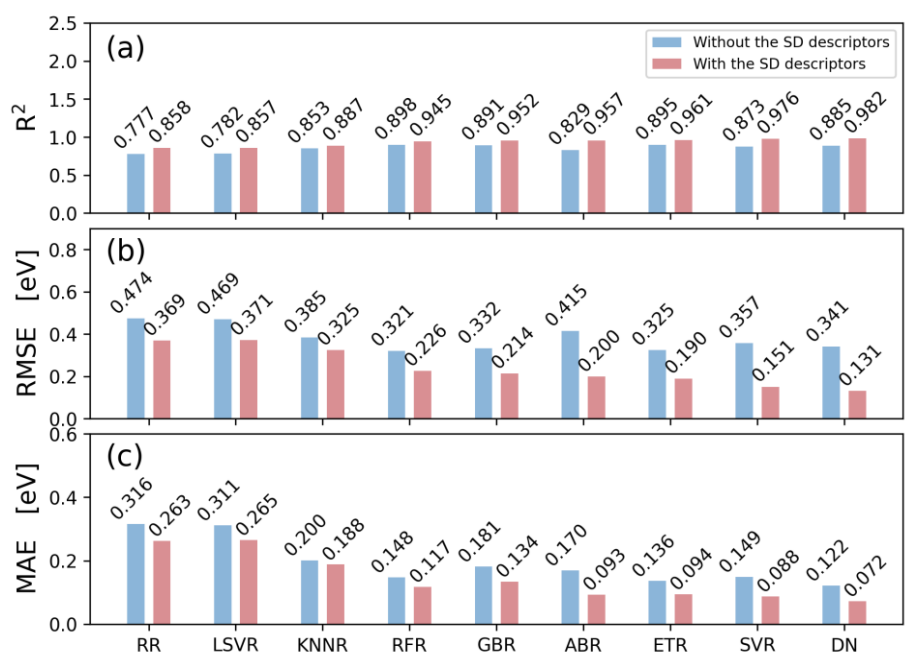

**Figure 3**. The (a) R2, (b) RMSE, and (c) MAE values of the nine different ML methods for formation energy prediction with and without the new proposed SD descriptors. The red and blue bars represent the corresponding values before and after adding these descriptors, respectively.

To analyze the control factors of $E_{form}$, the RFR method was chosen for ranking the importance of different descriptors, as that the RFR method is a tree-based learning algorithm and has advantages on both accuracy and interpretability [32]. Figure 4 shows the top 20 descriptors ranked for $E_{form}$ prediction, and the correlation between descriptor pair in the way of Pearson correlation coefficient matrix. It is found that the Pauling electronegativity is the most important factor for the prediction model of $E_{form}$ without and with SD descriptors (47.1% of $Pa\ \chi^r$ in Figure 4a and 46.4% of $\Delta Pa\ \chi^a$ in Figure 4b, respectively). Although the ratio of $\Delta Pa\ \chi^a$ slightly dropped when using the SD descriptors, the overall ratio of the top 20 important descriptors increased from 74.6% to 82.6%. In Figure 4a and 4b, there are five descriptors [$Pa\ \chi^r$, $Pa\ \chi^k$, $AR\ \chi^k$, $AR\ \chi^s$, $Mu\ \chi^k$] related to electronegativity and two descriptors [$\Delta Pa\ \chi^a$, $\Delta AR\ \chi^a$] related to electronegativity difference appeared in the top 20 important descriptors, respectively. Metallic valence represents the oxidation state of the metal in alloy similarly to that of covalent compound. In the previous study for predicting the thermodynamic stability of perovskites, the most common oxidation state is one of the most important. The four CD descriptors [$MV^s$, $MV^v$, $MV^r$, $MV^k$] and three CD descriptors [$MV^s$, $MV^v$, $MV^r$] related to metallic valence [33, 34] appeared in the 20 important descriptors in both Figure 4a and 4b, respectively. Figure 4b illustrated that the density ($\rho$), coordination number ($CN^a$), band gap ($E_g$) are also important descriptors for improving the prediction performance of $E_{form}$ which is consistent with the results of others [13, 35, 36]. In the prior study of Ward *et al.* [3], it is pointed out that the formation energy of intermetallic compounds is best described by the variances in the melting point ($MP^v$) and number of *d* electrons between constituent elements ($V\_d^v$). The similar result is also observed in this study because there existed linear relationship between the melting point and boiling point. As shown in Figure 4a and 4b, the variance of the boiling point ($BP^v$) as well as the variance of *d* electrons and unfilled *d* electrons [$V\_d^v$, $UV\_d^v$] appeared in the top 20 important descriptors. The variance of cohesive energy, the skewness and kurtosis of the 1st ionization potential [$CE^v$, $1^{st}\ IP^s$, $1^{st}\ IP^k$], also appears in the top 20 important descriptors.

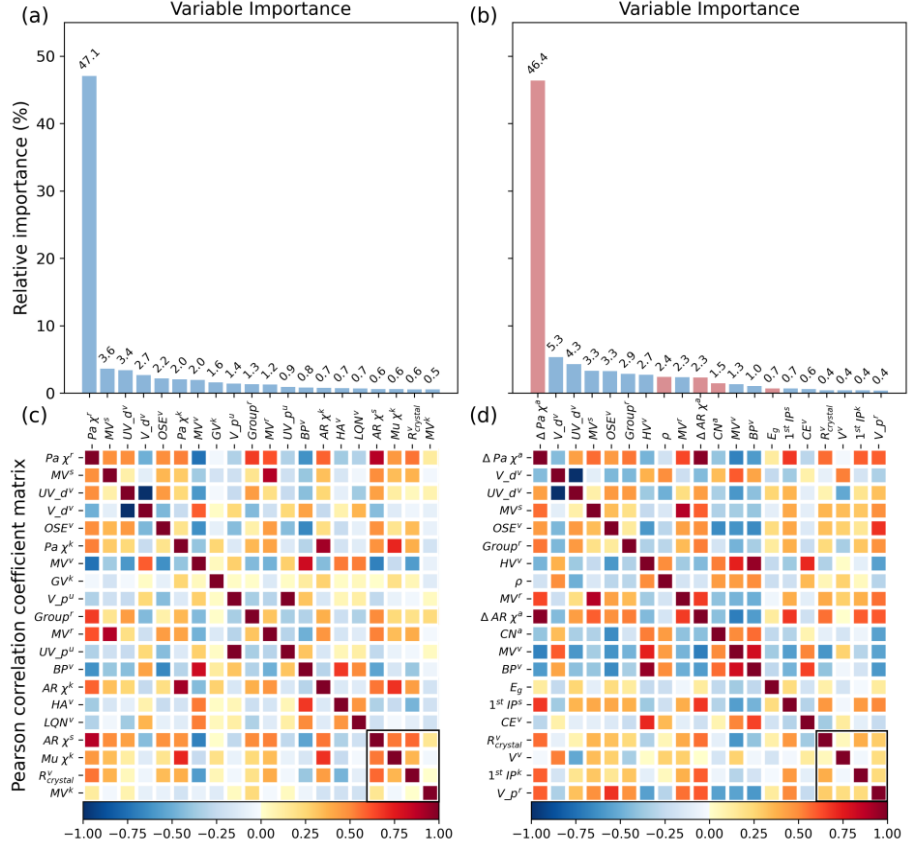

**Figure 4**. The importance ranking of all descriptors (a) without and (b) with the new SD descriptors in predicting $E_{form}$. The bars marked by the red color in (b) represent the relative importance of the SD descriptors proposed in this study. Pearson correlation coefficient matrix of (c) without and (d) with the SD descriptors. The lower-case letters a, v, r, s, k, and u appeared in the upper-right corner of x- and y-axis labels are the abbreviation of "average, variance, range, skewness, kurtosis, and sum".

Figure 4c and 4d show the Pearson correlation matrix for the top 20 important descriptors without and with the SD descriptor set. For each pair of descriptors, $A$ and $B$, the correlation coefficient $R_{AB}$ is defined as

$$R_{AB} = \frac{\sum_{i=1}^{n}(A_i - \bar{A})(B_i - \bar{B})}{\sqrt{\sum_{i=1}^{n}(A_i - \bar{A}))}\sqrt{\sum_{i=1}^{n}(B_i - \bar{B})}} \quad (10)$$

where $\bar{A}$ and $\bar{B}$ are the sample means of $A$ and $B$ descriptors over the total $n$ crystal materials, while $A_i$ and $B_i$ are the descriptors of the $i^{th}$ material. As shown in Figure 4c, strong correlations are found in several pairs of descriptors: (1) two descriptors related to the $d$ valence electrons orbitals [$UV\_d^v$, $V\_d^v$]; (2) three descriptor related to the electronegativity [$Pa\ \chi^k$, $AR\ \chi^k$, $Mu^k$]; (3) two descriptors related to the thermodynamic property [$BP^v$, $HA^v$]. Compared Figure 4c to 4d, the changes in the color distributions at the lower right corner (marked by the black rectangles) show that the correlations of the top 20 descriptors decreased obviously by adding the SD descriptors. Our results show that there is a correlation between different definitions of electronegativity. As shown in Figure S2 and S3, when the most important descriptor $\Delta Pa\ \chi^a$ in

Figure 4b was removed from the collection of the input descriptors, the $\Delta AR\ \chi^a$ descriptor became the most important descriptor after ranking again by the RFR method. Repeatedly the above steps, when removing $\Delta AR\ \chi^a$, the $\Delta AR\ \chi^u$ became the most important descriptor, which is consistent with the observed result in Figure 4d that there exists strong correlations among the descriptors, [$\Delta Pa\ \chi^a$, $\Delta AR\ \chi^a$ and $\Delta AR\ \chi^u$] and the relation on prediction performance of $E_{form}$ $\Delta Pa\ \chi^a$ > $\Delta AR\ \chi^a$ > $\Delta AR\ \chi^u$.

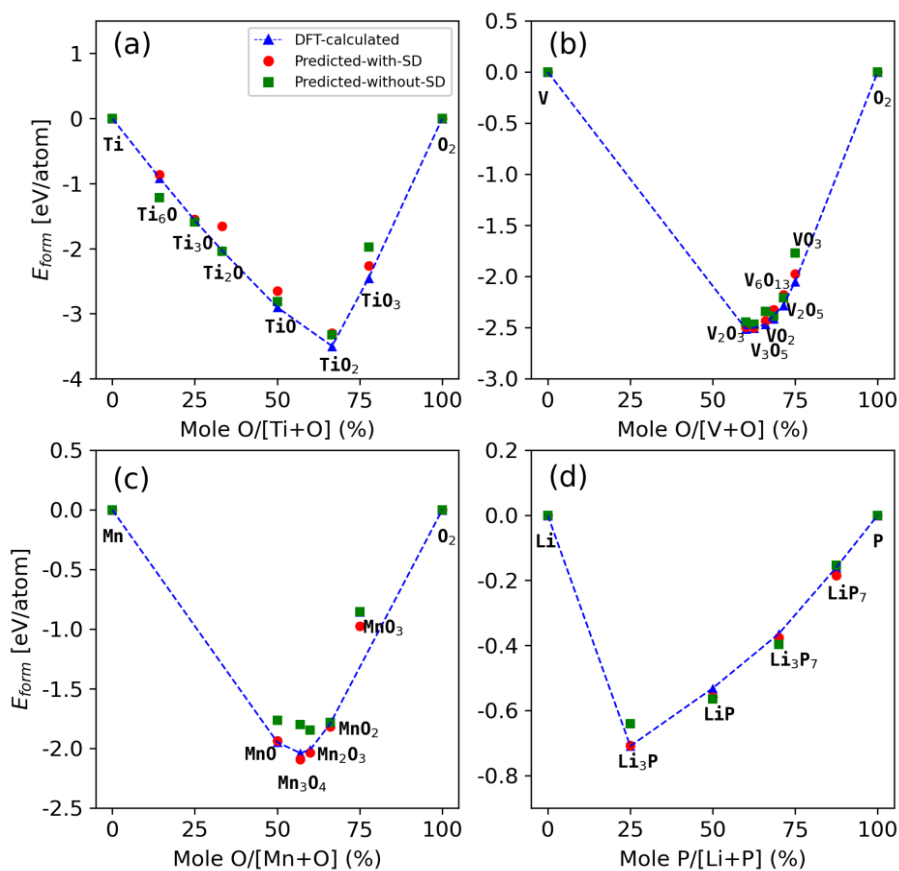

**Figure 5**. The thermodynamic phase diagram for compounds consisted in (a) Ti-O, (b) V-O, (c) Mn-O, and (d) Li-P chemical systems. The blue triangles, red circles and green rectangles indicate the DFT-calculated $E_{form}$s, predicted ones with and without the SD descriptors proposed in this study, respectively. The blue envelopes are made up of the DFT-calculated $E_{form}$s of the stable structures in each chemical system.

Figure 5 shows the thermodynamic phase diagram for compounds in Ti-O, V-O, Mn-O and Li-P chemical systems obtained from the DN model and DFT. These four chemical systems are selected to demonstrate the accuracy of the model for their importance as $TiO_x$ is used in photovoltaic devices to improve device preformation [37, 38], $VO_x$ is a promising cathode material and likely to be commercialized applications in the future [39], $MnO_x$ is as widely studied as the catalysts [40-42], and $LiP_x$ is the intermedia product in certain anode reactions of Li-ion batteries [43]. The thermodynamic phase diagram are generated use the same energy correction mechanism as the Materials Project [1, 44], to make sure the energy hulls in this work are comparable to either Materials Project [1, 44] or Atomly [19] (also adopts the same energy correction for generating the phase diagram). As shown in Figure 5a-5d, The DFT data tells us that there are 6 stable crystalline compounds [$Ti_6O$, $Ti_3O$, $Ti_2O$, $TiO$, $TiO_2$ and $TiO_3$] in the Ti-O chemical system, 4 stable crystalline compounds [$V_2O_3$, $V_3O_5$, VO and $VO_2$] and 2 unstable ones [$V_6O_{13}$ and $VO_3$] in the V-O chemical system, 4 stable crystalline compounds [MnO, $Mn_3O_4$, $Mn_2O_3$ and $MnO_2$] and 1 unstable one [$MnO_3$] in the Mn-O chemical system, and 4 stable crystalline compounds [$Li_3P$, LiP, $Li_3P_7$ and $LiP_7$] in the Li-P chemical system, respectively. Note that the DN model can assess the formation energy of those compounds fairly well. It is seen from Figure 5c that in Mn-O chemical system, there is a distinct advantage of the model with the SD descriptors on predictive performance over that without the SD descriptors. The other sub figures in Figure 5 show that both models, with or without the SD descriptors, can capture the shape of energy hull to some extent. When the SD descriptors is taken account into the model, the predicted $E_{form}$ is off only by ~135 meV·atom$^{-1}$ averagely (ranging from 0 to 373 meV·atom$^{-1}$) for Ti-O chemical system, ~40 meV·atom$^{-1}$ averagely (ranging from 0 to 102 meV·atom$^{-1}$) for V-O chemical system, ~33 meV·atom$^{-1}$ (ranging from 0 to 120 meV·atom$^{-1}$) for Mn-O chemical system, and ~11 meV·atom$^{-1}$ (ranging from 0 to 27 meV·atom$^{-1}$ ) for Li-P chemical system comparing with the DFT value. Hence, our model makes formation prediction easy and accurate.

## 3. Conclusions

In summary, we proposed several new structure-dependent descriptors those related to electronegativity difference and coordination numbers for predicting formation energy of a crystal material deriving from the local environment of the material via the Voronoi tessellation method. We demonstrated that these new descriptors can significantly improve the prediction accuracy of formation energy not only for eight classical ML methods but also for one neural network method. The neural network model achieved the highest prediction accuracy of $R^2$ = 0.982. The tree-based RFR method was chosen for screening the key descriptors for describing formation energy. It is found that the Pauling electronegativity difference between bonding atoms is the most important factor with a ratio of 46.4% for the prediction of formation energy. Our work shows that by adopting the physics-based descriptor as well as a good dataset, the predictive power of the machine learning model can be significantly improved. There might be still a large room out there to keep enhancing the predictive power if we can find better descriptors with better data. It is our hope that the model can provide a useful tool for the materials science community.

## 4. Methods

Eight conventional ML models as well as a neural network model were adopted for predicting the DFT-calculated formation energy per atom ($E_{form}$), which are AdaBoost Regression (ABR), Linear Support Vector Regression (LSVR), Support Vector Regression (SVR), Ridge Regression (RR), GradientBoosting Regression (GBR), K-Nearest Neighbors Regression (KNNR), Random Forest Regression (RFR), ExtraTrees Regression (ETR), and DenseNet (DN) [45, 46]. In this study, the commonly used ML models are implemented by *scikit-learn* [47] and the DenseNet is implemented by *pytorch* [48]. DenseNet is a neural network model, which has three hidden dimensions with node counts of [256, 128, 64], in addition to an input layer with 119 nodes and an output layer with 1 node. The DenseNet model was trained using the Mean Absolute Error (MAE) loss criterion, or called L1 loss criterion, shown in formula (1) and an Adam optimizer with a learning rate of 10$^{-3}$.

$$L1 = |f(x) - Y| \qquad (1)$$

where $f(x)$ represents the predicted value and $Y$ represents the actual DFT-calculated value retrieved from the Atomly materials-data-infrastructure. The predictive performances of models are described as variance ($R^2$), Mean Absolute Error (MAE), Root Mean Squared Error (RMSE).

## 5. Data availability

The dataset used in this work is obtained from the "atomly.net" [19], which is a DFT materials-data-infrastructure conceptually similar to the Materials Project, AFLOWlib and OQMD [4]. The data is generated high-throughputly with an automated workflow to crunch through more than 170,000 inorganic crystal materials. The dataset contains the crystal structure (*e.g.*, crystal structure, X-ray diffraction (XRD)), electronic structure (*e.g.*, density of states (DOS) and band gap ($E_g$)) and energies (*e.g.*, formation energy ($E_{form}$), energy above the convex hull ($E_{hull}$)) of the compounds, basically the properties those can be obtained from DFT. The data is calculated by using the "Vienna Ab Initio Simulation Package" (VASP) [49-51], and the Perdew-Burke-Ernzerhof (PBE) [52] exchange-correlation function of generalized gradient approximation (GGA) was used to deal with the interactions between electrons. The cutoff energy is 520eV and the k-mesh are 3000/Å$^3$.

The dataset contains 170,714 inorganic crystal materials in total as the time of writing, out of which there are 2,205 elementary substances, 21,969 binary compounds, 70,760 ternary compounds, and 75,780 compounds with between 4 and 9 constituent elements. The dataset is randomly split into training and test sets using a ten-fold cross validation with their unique composition and formation energy.

## 6. Acknowledgments

The computational resource is provided by the Platform for Data-Driven Computational Materials Discovery of the Songshan Lake laboratory. We especially thank Atomly database for data sharing. We would also acknowledge the financial support from the Research Program of the Basic Frontier Sciences for Original Innovation from 0 to 1, CAS (No. ZDBS-LY-SLH007), and Strategic Leading Science and Technology Project, CAS (No. XDB33020000).